\documentclass[11pt]{article}
\usepackage{moriond,epsfig}

\def\Journal#1#2#3#4{{#1} {\bf #2}, #3 (#4)}

\def\NIM{\em Nucl. Instrum. Methods}

\def\NPA{{\em Nucl. Phys.} A}

\def\PRL{\em Phys. Rev. Lett.}

\def\PRC{{\em Phys. Rev.} C}

\def\be{\begin{equation}}
\def\ee{\end{equation}}
\def\bea{\begin{eqnarray}}
\def\eea{\end{eqnarray}}

\begin{document}
\vspace*{4cm}
\title{LATEST RESULTS FROM PHOBOS AT RHIC}

\author{G\'abor I. VERES$^{4,*}$ for the PHOBOS Collaboration\\
\vspace{2mm}
B.B.Back$^1$,
M.D.Baker$^2$,
M.Ballintijn$^4$,
D.S.Barton$^2$,
R.R.Betts$^6$,
A.A.Bickley$^7$,
R.Bindel$^7$,
W.Busza$^4$,
A.Carroll$^2$,
Z.Chai$^2$,
M.P.Decowski$^4$,
E.Garc\'{\i}a$^6$,
T.Gburek$^3$,
N.George$^2$,
K.Gulbrandsen$^4$,
C.Halliwell$^6$,
J.Hamblen$^8$,
M.Hauer$^2$,
C.Henderson$^4$,
D.J.Hofman$^6$,
R.S.Hollis$^6$,
R.Ho\l y\'{n}ski$^3$,
B.Holzman$^2$,
A.Iordanova$^6$,
E.Johnson$^8$,
J.L.Kane$^4$,
N.Khan$^8$,
P.Kulinich$^4$,
C.M.Kuo$^5$,
W.T.Lin$^5$,
S.Manly$^8$,
A.C.Mignerey$^7$,
R.Nouicer$^{2,6}$,
A.Olszewski$^3$,
R.Pak$^2$,
C.Reed$^4$,
C.Roland$^4$,
G.Roland$^4$,
J.Sagerer$^6$,
H.Seals$^2$,
I.Sedykh$^2$,
C.E.Smith$^6$,
M.A.Stankiewicz$^2$,
P.Steinberg$^2$,
G.S.F.Stephans$^4$,
A.Sukhanov$^2$,
M.B.Tonjes$^7$,
A.Trzupek$^3$,
C.Vale$^4$,
G.J.van~Nieuwenhuizen$^4$,
S.S.Vaurynovich$^4$,
R.Verdier$^4$,
E.Wenger$^4$,
F.L.H.Wolfs$^8$,
B.Wosiek$^3$,
K.Wo\'{z}niak$^3$,
B.Wys\l ouch$^4$\\
\vspace{3mm}
\small
$^1$~Argonne National Laboratory, Argonne, IL 60439-4843, USA\\
$^2$~Brookhaven National Laboratory, Upton, NY 11973-5000, USA\\
$^3$~Institute of Nuclear Physics PAN, Krak\'{o}w, Poland\\
$^4$~Massachusetts Institute of Technology, Cambridge, MA 02139-4307, USA\\
$^5$~National Central University, Chung-Li, Taiwan\\
$^6$~University of Illinois at Chicago, Chicago, IL 60607-7059, USA\\
$^7$~University of Maryland, College Park, MD 20742, USA\\
$^8$~University of Rochester, Rochester, NY 14627, USA\\}

\address{$^*$present address: Department of Atomic Physics, E\"otv\"os 
Lor\'and University\\
P\'azm\'any P\'eter s\'et\'any 1/A, Budapest H-1117, Hungary}

\maketitle

\abstracts{
A study of charged hadron production in d+Au and Au+Au collisions is 
presented at various 
collision energies ($\sqrt{s_{_{NN}}}$=19.6 to 200 GeV). Scaling and 
factorization features of $p_T$ and $\eta$ distributions and $v_2$ 
are discussed as a function of $\sqrt{s_{_{NN}}}$ and collision 
centrality.}

\section{Introduction}

\begin{figure}
\centerline{\psfig{figure=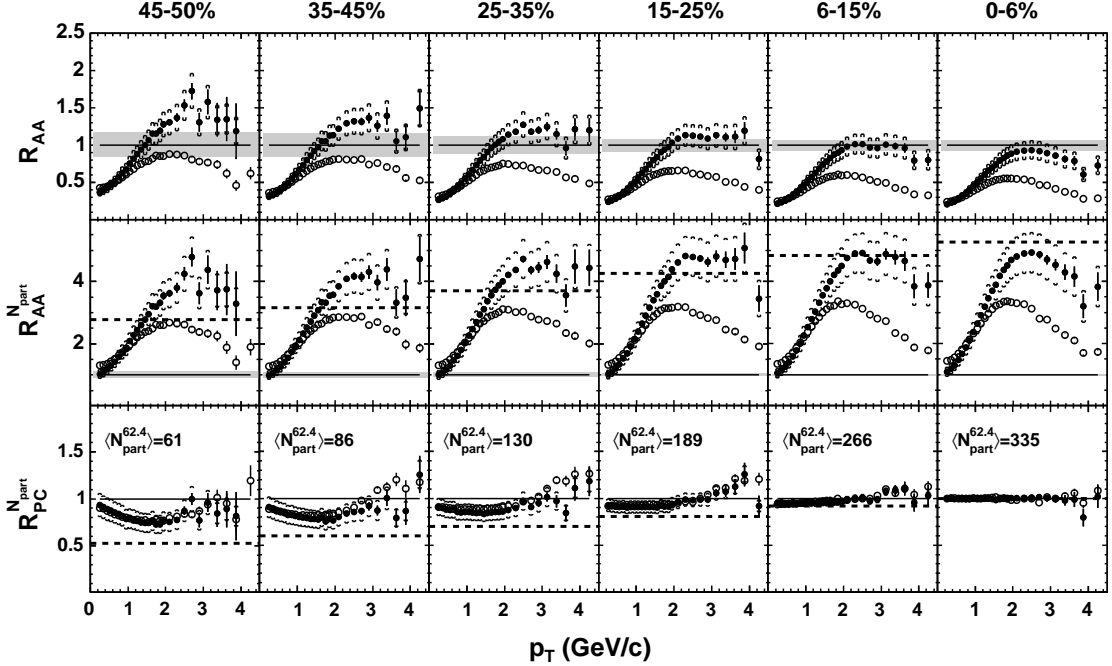,width=150mm}}
\caption{Ratio of $p_T$ distributions from Au+Au collisions to various
reference distributions at $\sqrt{s_{_{NN}}}$=62.4 GeV (filled symbols)
and 200 GeV (open symbols), in six centrality bins. The $R_{AA}$,
$R_{AA}^{N_{part}}$, $R_{PC}^{N_{part}}$ quantities are discussed in the
text. The dashed line is the $N_{coll}$ scaling expectation; the gray
band is the uncertainty of $N_{coll}$.
\label{master}}
\end{figure}

The study of heavy ion collisions constitutes an important part of the recent 
experimental and theoretical effort to understand the strong interaction 
which binds quarks and gluons into hadrons.
Predictions based on QCD indicate the existence of a new state of matter dominated by the 
strong interaction, where bound hadrons no longer 
exist, provided the energy density is sufficiently high (greater than a 
GeV/fm$^3$). Heavy ion collisions are the only method to create such a 
high energy density in the laboratory.
The current understanding of the phase structure of strongly 
interacting matter, the properties of the matter in the various phases and 
the nature of the transition between them is, to a large extent, driven by 
experiment. One of the important discoveries at the Relativistic Heavy Ion 
Collider (RHIC) is that an extremely high energy density system is 
created, where hadronic degrees of freedom cannot be relevant any more.
There is evidence for a very significant level of interaction between the 
constituents of this system, as opposed to earlier expectations. 

In this experimental talk we will emphasize a few basic scaling and 
factorization features of the data collected by the PHOBOS 
experiment at RHIC, in comparison 
with earlier measurements.
These simple rules highlight common features of collisions of heavy ions 
(Au+Au) and simpler systems (d+Au, p+p) in a broad range of collision 
energies ($\sqrt{s_{_{NN}}}$=6.7 to 200 GeV). Details of these findings 
can be found in a volume summarizing the results of the four experiments 
from the first three years of RHIC, in which our 
contribution~\cite{white} is titled `The PHOBOS Perspective on 
Discoveries at RHIC'.
To collect the data presented here, we used the magnetic spectrometer 
and the multiplicity arrays (covering the $|\eta|<5.4$ 
pseudo-rapidity region) of the PHOBOS 
experiment (described in detail elsewhere~\cite{nim}).

\section{Scaling features in heavy ion data}

We concentrate on three topics: transverse momentum ($p_T$) and 
pseudo-rapidity ($\eta$) distributions of charged hadrons, and the 
azimuthal asymmetry of their production which is called elliptic flow. 
The consensus in the heavy ion community is that in the early stage 
of high energy heavy ion collisions a new state of strongly interacting 
matter has been created. The above observables are related to relevant 
physical characteristics of this newly created matter, such as the
suppression of high-$p_T$ particles (jet quenching); the initial energy 
density and entropy after the collision and the boost-invariance of  
particle production along the colliding beam direction; and the collective 
motion of particles resulting from secondary interactions and the properties 
of the equation of state.


Determining the {\bf centrality} of a heavy ion collision is essential to 
provide the basis of comparison with more elementary (p+p, 
p+$\overline{\rm p}$) processes. Instead of the impact parameter, two 
different quantities are used to quantify the centrality: the number of 
participant nucleon pairs 
($N_{part}/2$) and the number of binary nucleon-nucleon collisions 
($N_{coll}$). Since the nucleon-nucleon cross section increases with 
collision energy, so does the $N_{coll}/N_{part}$ ratio, which also grows 
with decreasing impact parameter for simple geometrical reasons.
These two quantities are calculated by measuring the charged hadron 
multiplicity in various regions of $\eta$, combined with
a comprehensive Monte-Carlo simulation including the Glauber 
model.~\cite{richard}

If one normalizes the {\bf transverse momentum distribution} of charged 
hadrons measured in 
different centrality bins to the $p_T$ distribution measured in p+p($\overline{\rm p}$) 
collisions and also divides by $N_{coll}$, one observes a 
gradual decrease of this quantity ($R_{AA}$) with decreasing impact 
parameter (Fig. \ref{master}). Note that the expectation for `hard' 
collisions with large 
momentum transfer and no re-interaction with the created medium would be a
constant $R_{AA}$ equal to unity. However, the $R_{AA}^{N_{part}}$ 
quantity (where we replaced $N_{coll}$ by $N_{part}$ in the denominator)
scales with centrality much more precisely in Au+Au collisions at 
$\sqrt{s_{_{NN}}}$=62.4 and 200 GeV,~\cite{auau62spectra}
pointing to new physical interpretations.~\cite{muller} 
More strikingly, the $p_T$ spectra normalized to the most central bin at 
each energy, and also normalized by $N_{part}$, as illustrated in the 
bottom row of Fig. \ref{master}, are identical within 
errors in each of our centrality bins. This is a clear factorization of 
collision energy and centrality.
It will be interesting to compare these conclusions to the data from 
half a billion Cu+Cu events taken in the present RHIC run, 
since, for the same number of participants, the collision zone will have a 
very different geometry in the Cu+Cu and the Au+Au events.


\begin{figure}
\centerline{
\psfig{figure=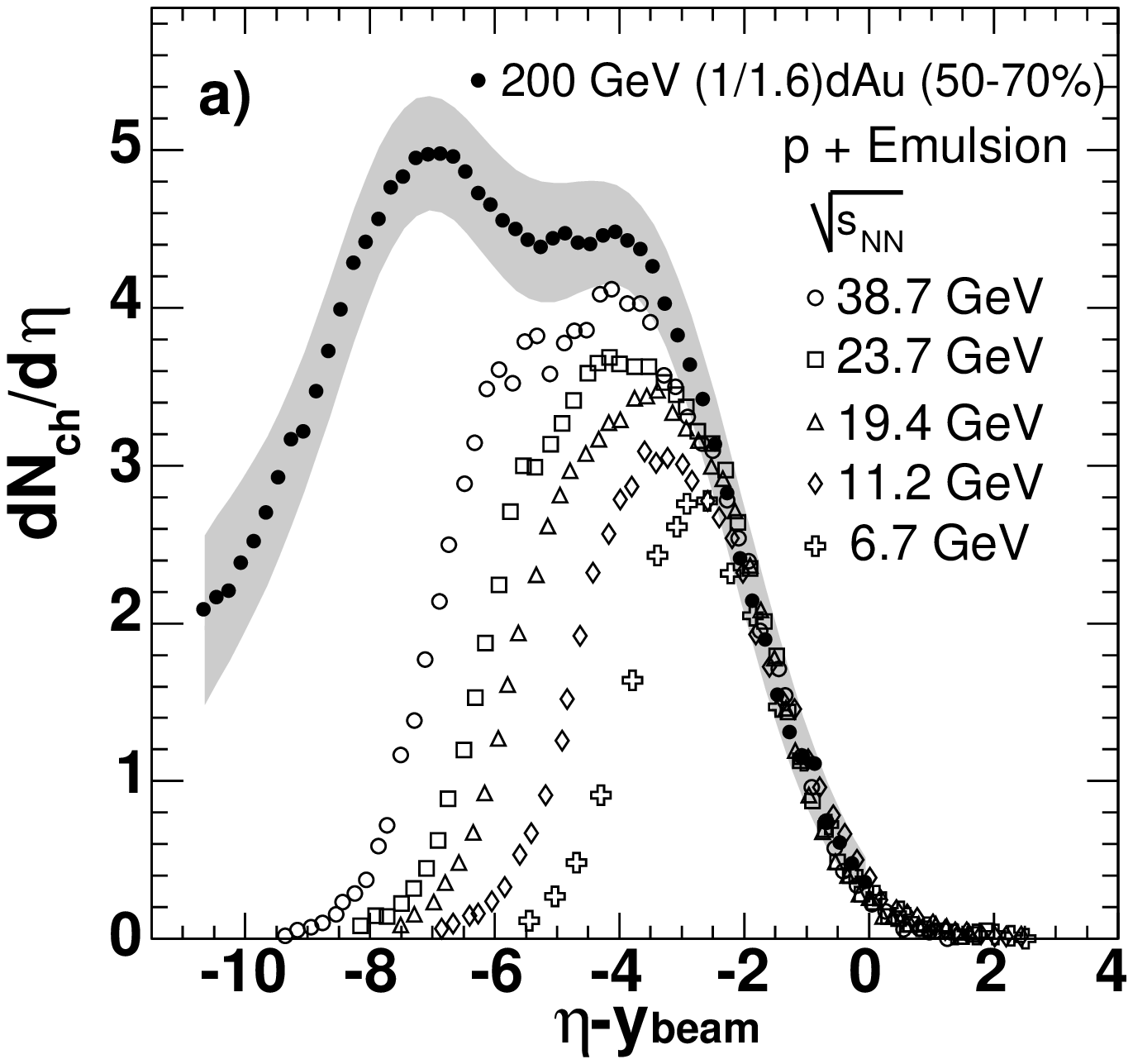,width=50mm}
\psfig{figure=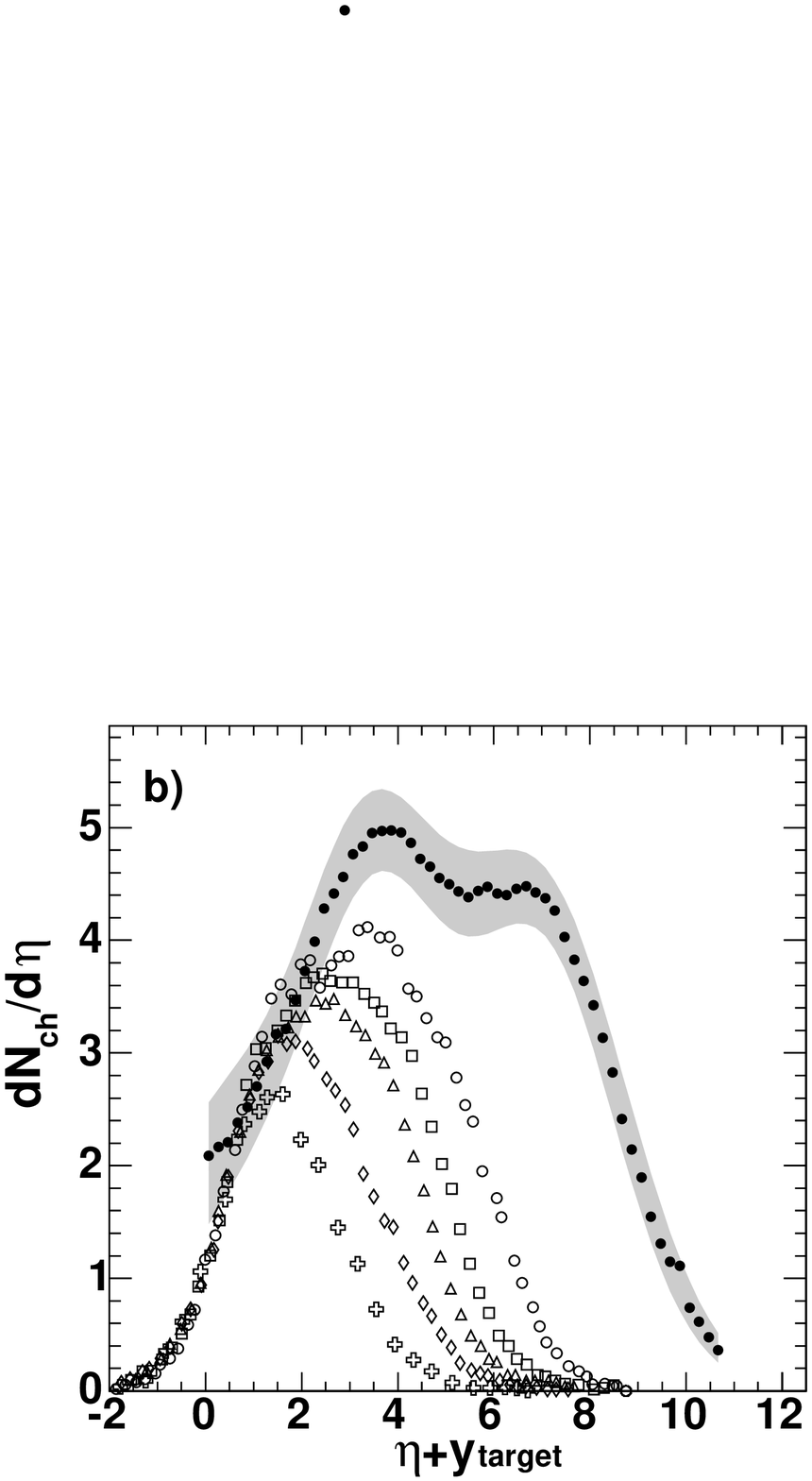,width=50mm}
\psfig{figure=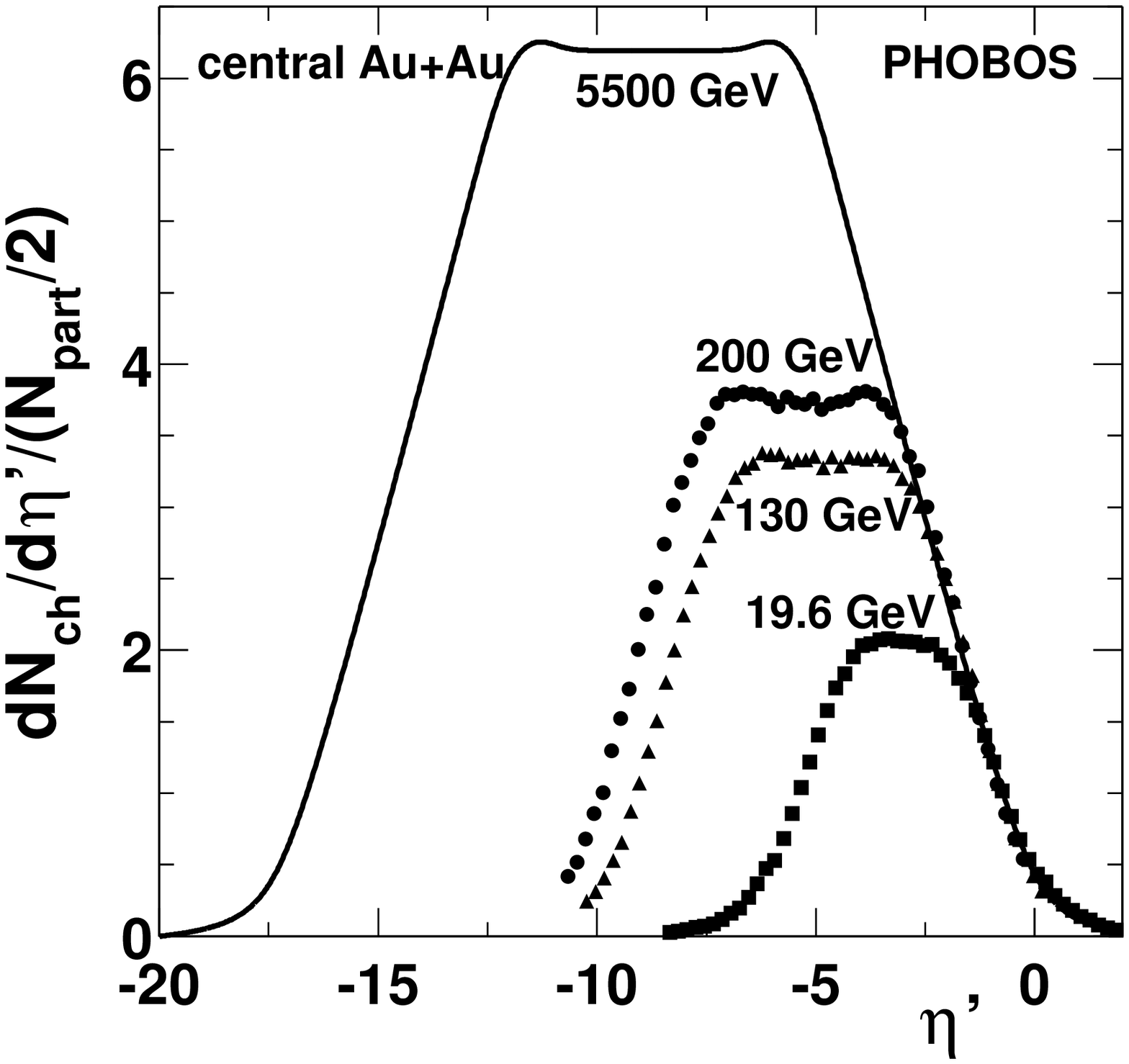,width=49.5mm}}
\caption{Pseudo-rapidity density distributions of charged particles 
measured in p+Emulsion and d+Au collisions at various energies,
approximately shifted to the rest frame of the lighter (a) and heavier (b) 
species. On the rightmost panel, scaling features of Au+Au data are used 
to predict the $\eta'$ distribution at the LHC (Pb+Pb).
\label{mult}}
\end{figure}

Another simple feature of the data is the {\bf extended longitudinal 
scaling of pseudo-rapidity} ($\eta$) {\bf density} distributions of charged 
particles, in p+Emulsion and d+Au,~\cite{white} as shown in 
Fig. \ref{mult}. By plotting these distributions as a function of 
$\eta'=\eta\pm y_{target}$, thus effectively looking at them from the rest 
frame of one of the colliding beams, we observe that the data at various 
energies fall on a common limiting curve, in both reference frames.
The longitudinal scaling extends
over more than an order of magnitude in beam energy. 

A similar scaling was observed earlier in heavy ion (Au+Au) collisions
by PHOBOS, illustrated on the right panel of Fig. \ref{mult} for the 6\% 
most central data. In addition, applying the fact that $dN/d\eta$
at $\eta\approx 0$ in central collisions scales logarithmically with 
$\sqrt{s_{_{NN}}}$, we can extrapo\-late the $dN/d\eta$ distribution 
to LHC energies and give a simple experiment-based prediction~at 
$\sqrt{s_{_{NN}}}$=5500 GeV. The prediction gives 
$dN/d\eta|_{\eta=0}\approx$ 1100 and 14000 charged particles in 
total.~\cite{wit}



\begin{figure}
\centerline{\psfig{figure=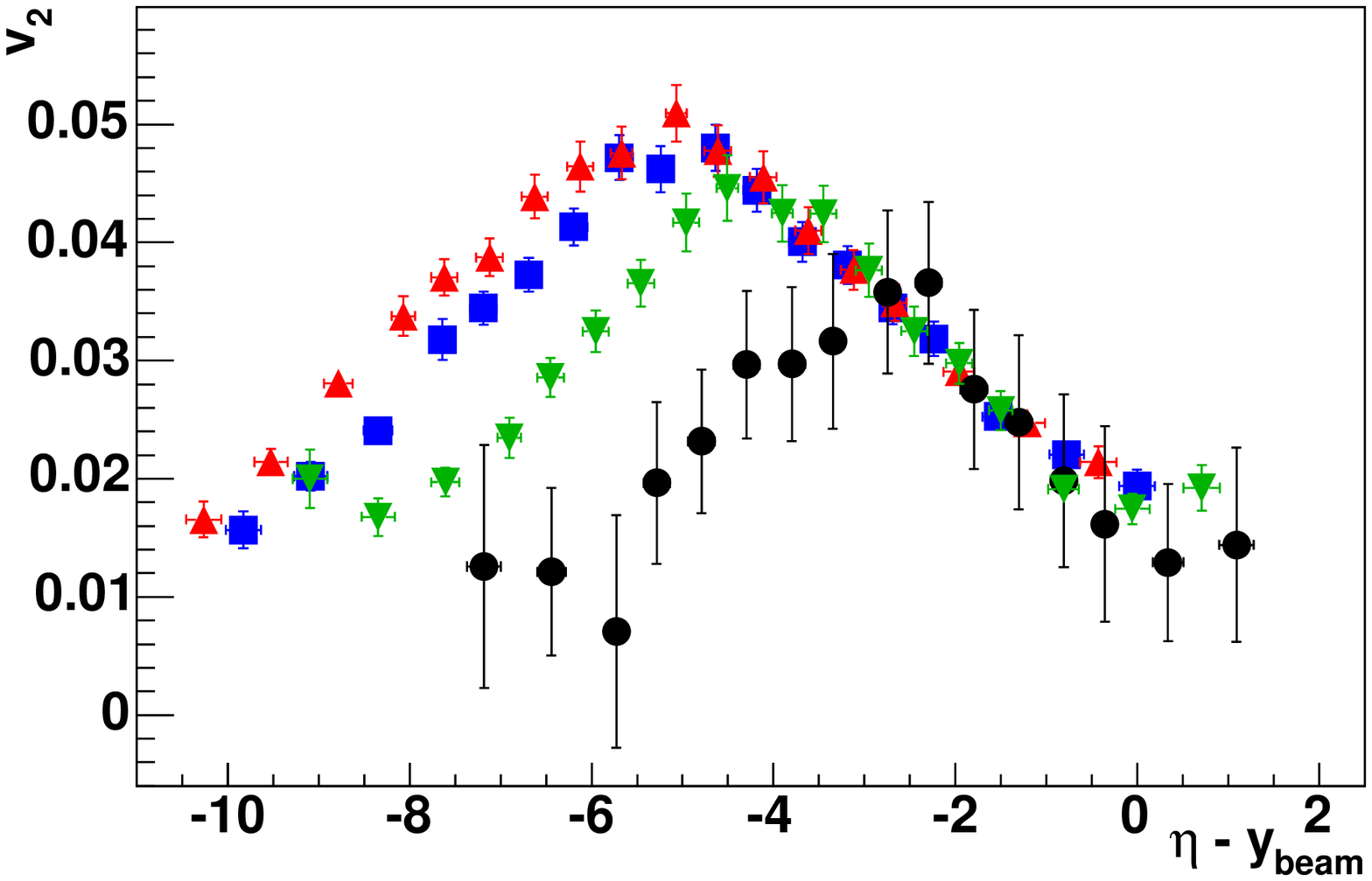,width=78mm}
\psfig{figure=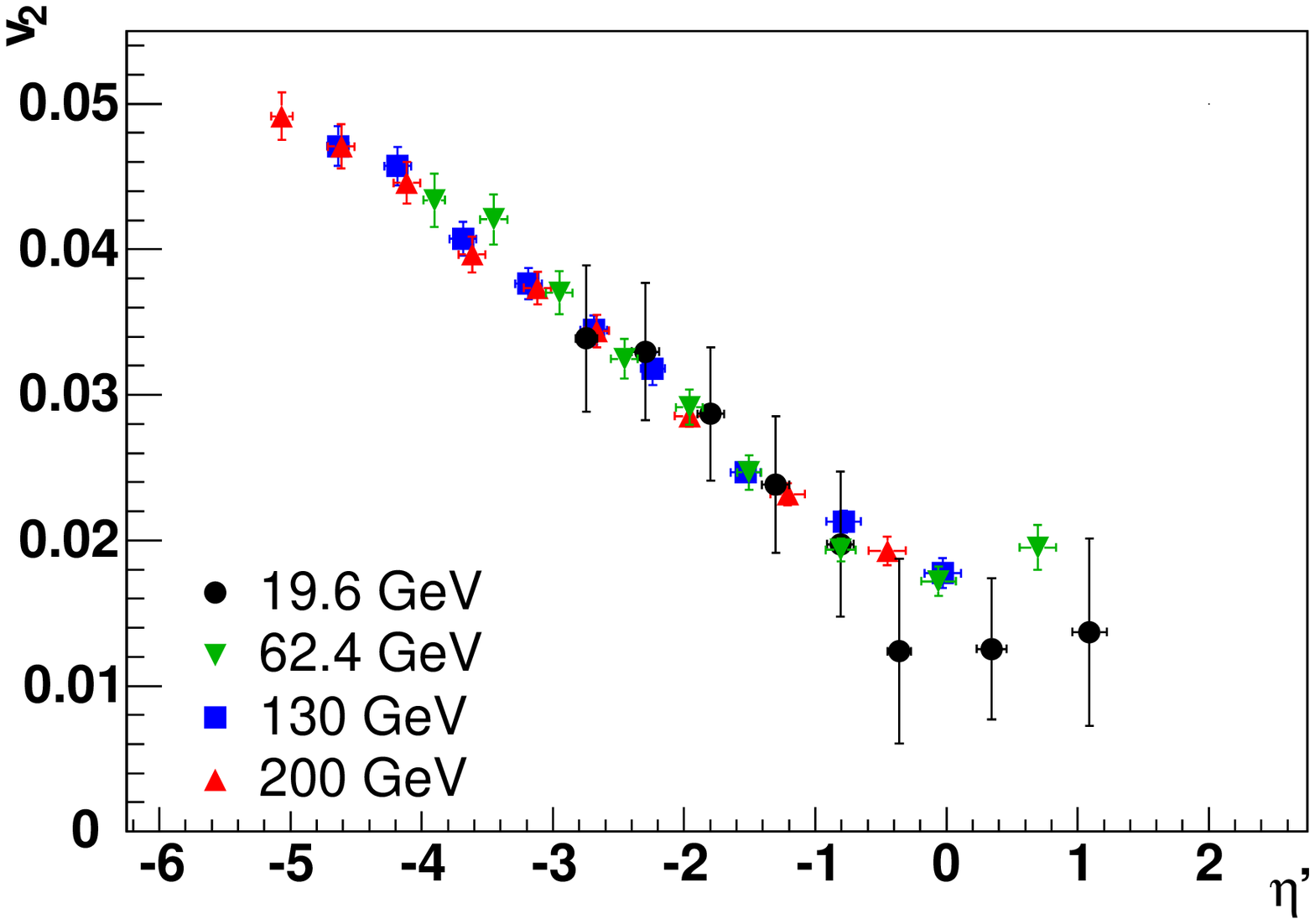,width=78mm}}
\caption{
The elliptic flow parameter $v_2$, averaged over centrality (0-40\%), as a 
function of 
$\eta'=\eta-y_{beam}$ (left) and $|\eta|-y_{beam}$ (right) for various 
collision energies (Au+Au). Only statistical error bars are shown.
\label{flowfig}}
\end{figure}

Longitudinal scaling has also been recently observed in the 
{\bf elliptic flow} of particles produced in heavy ion 
collisions.~\cite{flow}
The elliptic flow parameter ($v_2$) is a sensitive probe of the properties of the 
newly created, very dense and hot matter in 
the early stage of the collision. In Fig. \ref{flowfig}, the 
pseudo-rapidity dependence of this $v_2$ parameter is plotted for 
semi-central Au+Au events and for various energies, where we use the 
$\eta'=\eta-y_{beam}$ parameter again. There seems to be a 
universal curve governing $v_2$ over a broad range of 
$\eta'$. This is shown more precisely on the right panel of Fig. 
\ref{flowfig}, where we used the symmetry and plotted $v_2$ as a function 
of $|\eta|-y_{beam}$. 

\section{Conclusions}

We reported a few simple scaling and factorization properties of charged 
hadron production in heavy ion collisions at RHIC energies. 
We observed that the transverse momentum spectra approximately scale
with the number of participant nucleons, and the energy and centrality 
dependence of these spectra precisely factorize in the range of centrality 
and $p_T$ we studied. Longitudinal scaling was seen in an 
extended pseudo-rapidity region in Au+Au (and d(p)+A) collisions,
both in case of the $dN/d\eta$ pseudo-rapidity density and in case of the 
$v_2$ elliptic flow parameter. These simple features of the data impose 
constraints on models attempting to describe and understand the basic 
particle production mechanism in heavy ion collisions.

\section*{Acknowledgments}
This work was partially supported by U.S. DOE grants 
DE-AC02-98CH10886,
DE-FG02-93ER40802, 
DE-FC02-94ER40818,  
DE-FG02-94ER40865, 
DE-FG02-99ER41099, and
W-31-109-ENG-38, by U.S. 
NSF grants 9603486, 
0072204,            
and 0245011,        
by Polish KBN grant 1-P03B-062-27(2004-2007),
by NSC of Taiwan Contract NSC 89-2112-M-008-024, and
by the Hungarian Scientific Research Fund (OTKA F049823) and the J\'anos Bolyai 
Research Grant.

\end{document}